\documentclass[aps,prl,superscriptaddress,floatfix,twocolumn]{revtex4-2}
\usepackage{graphics}
\usepackage{graphicx}
\usepackage{subfigure}
\usepackage{mathrsfs}
\usepackage{amsmath}
\usepackage{color}
\usepackage{natbib}
\usepackage{bookmark}

\begin{document}

\title{Anomalous Transport of Elongated Particles in Oscillatory Vortical Flows}

\author{Shiyuan Hu}
\thanks{These authors contributed equally to this work.}
\email[]{shiyuanhu@buaa.edu.cn}
\affiliation{School of Physics, Beihang Univerity, Beijing 100191 China}
\affiliation{CAS Key Laboratory of Theoretical Physics, Institute of Theoretical Physics, Chinese Academy of Sciences, Beijing 100190, China}
\author{Xiuyuan Yang}
\thanks{These authors contributed equally to this work.}
\affiliation{School of Physics, Beihang Univerity, Beijing 100191 China}
\author{Nan Luo}
\affiliation{Beijing National Laboratory for Condensed Matter Physics and Laboratory of Soft Matter Physics, Institute of Physics, Chinese Academy of Sciences, Beijing 100190, China}
\affiliation{School of Physical Sciences, University of Chinese Academy of Sciences, Beijing 100049, China}
\author{Jun Zhang}
\email[]{jun@cims.nyu.edu}
\affiliation{Applied Mathematics Lab, Courant Institute of Mathematical Sciences, and Department of Physics, New York University, New York, NY 10012, USA}
\affiliation{NYU-ECNU Institute of Physics at NYU Shanghai, Shanghai 200062, China}
\author{Xingkun Man}
\email[]{manxk@buaa.edu.cn}
\affiliation{School of Physics, Beihang Univerity, Beijing 100191 China}
\affiliation{Peng Huanwu Collaborative Center for Research and Education, Beihang University, Beijing 100191, China}

\date{\today}

\begin{abstract}
We investigate the transport dynamics of elongated particles in cellular vortical flows that undergo spatial oscillations over time. Experimental flow visualizations reveal mixed flow fields with chaotic and elliptic regions coexisting. Surprisingly, the particle transport rate does not increase monotonically with particle length, even though longer particles are expected to explore neighboring vortices more easily. Numerical simulations in a much larger system produce similar transport anomalies, characterized by subdiffusion due to frequent long-time trapping in vortices at certain lengths, but normal diffusion at others. At moderate oscillation frequencies, these long-time trapping events occur within the chaotic region; at high frequencies, they occur in the elliptic regions, but only for particles whose lengths match these regions. In the latter case, subdiffusion is robust against random noise. Our results reveal new mechanisms for controlling particle diffusion in fluid flows.
\end{abstract}

\maketitle

The emergence and control of random walks of suspended particles in structured fluid environments have drawn considerable attention, with implications for target search~\cite{Volpe2017}, microbial dispersal in porous media~\cite{Bertrand2018,Mokhtari2019,Bhattacharjee2019,Kurzthaler2021}, and various chemical and biomedical processes such as particle sorting and mixing~\cite{Dorfman2013,Xuan2014,Khatami2016}. In steady flows at microscales where the Reynolds number (Re) is typically small, the emergence of such random and complex dynamics requires particles to deviate from flow streamlines. Diverse mechanisms, including particle motility~\cite{Marcos2012,Rusconi2014,Ariel2017,Brosseau2019}, inertial effects~\cite{Babiano00,Ouellette08}, deformability~\cite{Kantsler2012,Liu2018,Lagrone2019,Du2019,Bonacci2023,Hu2023}, and interparticle interactions~\cite{Pine05,Corte2008}, give rise to surprising phenomena. In particular, in two-dimensional cellular vortical flows---a canonical dynamical system for studying particle transport and advection~\cite{Leibovich1983,Solomon1988,Tabeling91,Rothstein99,Solomon2003,Bergougnoux2014,Lopez2017}---passive filaments display a rich variety of random walks and dispersal behaviors~\cite{Young07,Manikantan2013,Hu2021}, determined by their physical properties.

While previous studies reveal different microscopic mechanisms for cross-streamline dynamics, only steady, time-independent flows are considered. Time-dependent flows are ubiquitous in nature and are of growing interest in microfluidic applications~\cite{Guo2016,Dincau2020,Recktenwald2022,Mudugamuwa2024} and in the studies of active particle dynamics~\cite{Torney2007,Khurana2011,Ran2021,Qin2022}. Even at low Re, they allow fluid particles to depart from streamlines, leading to nontrivial dynamics. In such settings, however, the dynamics of passive, finite-length particles is much less explored. In two-dimensional flows, unsteadiness generally breaks integrability and produces a mixed, multiscale flow field: chaotic regions coexist with regular elliptic regions, in which fluid particles follow closed, quasiperiodic trajectories~\cite{Ottino1989}. Despite extensive research on chaotic advection of fluid in unsteady flows at low Re (see e.g.,~\cite{Aref2017,Speetjens2021}), it remains unclear how flow unsteadiness and underlying spatial scales affect the diffusion of advected elongated particles.

In this work, we combine experiments and numerical simulations to investigate the transport dynamics of elongated particles in time-dependent flows at low Re. Using electromagnetic forcing, we generate cellular vortical flows that oscillate spatially over time by shifting a magnet array. This differs from previous experiments in which time dependence is introduced by periodically reversing the standing vortices~\cite{Rothstein99,Voth02} or through fluid viscoelasticity~\cite{Liu2012}. We find that the particle transport depends on both the particle length and the oscillation frequency $f$ relative to the fluid advection time scale $\tau_0 = W/U_0$, where $W$ is the vortex size and $U_0$ is the maximum flow speed. Over a wide frequency range, we observe subdiffusion with frequent long-time trapping in vortices for particles of certain lengths, but normal diffusion at others. At moderate frequencies $f\sim 0.5 \ \tau_0^{-1}$, particles starting in the chaotic region are hardly affected by flows in the elliptic regions. However, at high frequencies $f\sim 1.0 \ \tau_0^{-1}$, they can penetrate the elliptic regions and become trapped inside, and only those whose lengths match these regions are subdiffusive. 

\begin{figure}[t]
\centering
\includegraphics[bb = 0 7 245 255, scale=1.0, draft=false]{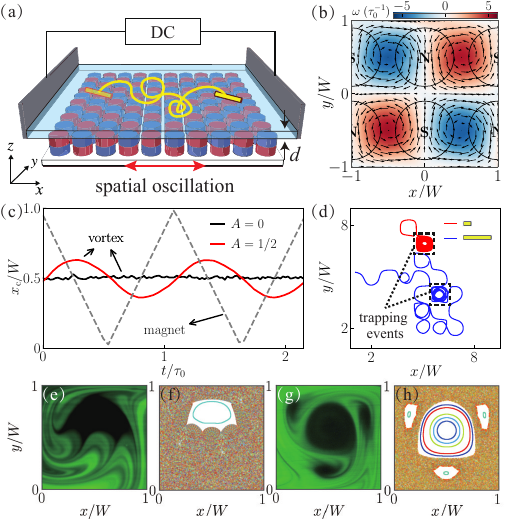}
\caption{Oscillatory cellular vortical flows and particle transport. (a) Experimental setup. (b) A periodic unit flow cell. Dark circles mark the static magnet positions without oscillation. Colormap shows flow vorticity. (c) Position of vortex center versus time. The dashed line shows the position of one magnet for $A = 1/2$. (d) Typical trajectories for short and long particles in steady flows. (e)--(h) Dynamical structures of the flow fields for $A_\mathrm{f} = 0.12$. Only one vortex is shown. In (e) and (f), $f=0.5$; in (g) and (h), $f=1.0$. (e),(g) Flow visualizations using fluorescent dye, taken after 10 oscillation periods. (f),(h) Poincar\'e sections of the model flow. Different colors represent trajectories from different initial conditions.}
\label{fig1}
\end{figure}
\textit{Experiment}---We generate quasi-two-dimensional cellular vortical flow fields using electromagnetic forcing~\cite{Tabeling91,Rothstein99}. A constant direct current (DC) is applied to a shallow layer of NaCl aqueous solution [Fig.~\ref{fig1}(a)]. By placing a $10 \times 10$ array of magnets with spatially alternating polarities beneath the fluid layer, an array of counter-rotating vortices forms via the Lorentz force. We measure the flow field in the upper fluid layer just below the free surface. As shown in Fig.~\ref{fig1}(b), an unit flow cell consists of four counter-rotating vortices. The vortex size $W=7$ mm, $U_0 \sim 2$ mm/s, and $\mathrm{Re} \sim 1$--10. Below, we scale all lengths by $W$, velocities by $U_0$, and time by $\tau_0 = W/U_0$. Fluid tracers in different vortices are observed to follow closed orbits and remain separated. To induce mixing, we program the magnet array to oscillate spatially with amplitude $A$ along a linear direction $\hat{\mathbf{x}}$. As a result, the vortex array oscillates spatially as a whole at the same frequency $f$ [Fig.~\ref{fig1}(c)].

Due to the rapid spatial decay of the magnetic field, the top fluid layer is mainly driven by lower layers through viscous stress. The velocity response across the fluid depth develops over a characteristic time $\tau_{\mathrm{d}} \sim d^2/\nu$, where $d$ is the fluid depth and $\nu$ is the kinematic viscosity, leading to a phase difference between the oscillations of the vortex and the magnet arrays [Fig.~\ref{fig1}(c)]. The flow oscillation amplitude $A_{\mathrm{f}}$ is smaller than $A$; by varying $A$ and $f$, we find that $A_{\mathrm{f}} \sim A \exp [-(f\tau_{\mathrm{d}}/\tau_0)^{1/2}]$ (see Supplemental Materials~\cite{SM}), which implies boundary-driven flows~\cite{Batchelor00}. Therefore, given $A$, $f$, and $d$, the flow field is uniquely determined. Since $A_{\mathrm{f}}$ decays rapidly with $f$ for a given $A$, we focus on the range $f\lesssim 1.0$ in this work.

With the imposed oscillation, fluid tracers in the regions between neighboring vortices can now move across them, exhibiting chaotic motion. To visualize the dynamical structures of the flow fields, we place fluorescent dye in a narrow band along the edges of each vortex and record its advection by the flow. The dye is quickly stretched and deformed but cannot penetrate certain regions (see video in~\cite{SM}). After several oscillation periods, persistent patterns are reached and recur once per period. Two snapshots are shown in Figs.~\ref{fig1}(e) and \ref{fig1}(g) for $f= 0.5$ and $f=1.0$, respectively. These regions inaccessible to advection are transport barriers for fluid particles, known as KAM islands~\cite{Ottino1989}. 
\begin{figure}[b]
\centering
\includegraphics[bb = 0 10 245 85, scale=1.0, draft=false]{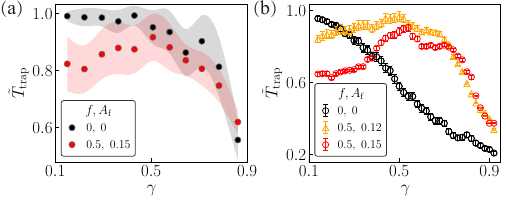}
\caption{$\tilde{T}_{\mathrm{trap}}$ as a function of $\gamma$ obtained from (a) experiments and (b) simulations. Particles are initially randomly distributed in a unit cell. Shaded areas and the error bars represent 1 standard deviation.}
\label{fig2}
\end{figure}

With the understanding of flow structures, we investigate the dynamics of elongated, ABS plastic particles with lengths $\gamma = 0.1$--1.0. Due to density difference, they float at the free surface. For comparison, we first examine the case without oscillation. For each $\gamma$, $N=50$ trajectories are tracked with random initial positions, each lasting for a duration $T\approx 150\ \tau_0$. Short particles ($\gamma \ll 1$) are seen to behave like fluid tracers and are typically trapped in vortices for extended periods of time, whereas long particles ($\gamma \sim 1$) usually move across different vortices before getting to the boundaries of the container [Fig.~\ref{fig1}(d) and video in~\cite{SM}]. A trapping event is defined as one in which a particle remains inside a vortex for at least two full circulations. The overall transport rate of particles of a given length is inversely related to the total trapping time $T_{\mathrm{trap}}$, obtained by summing the time durations of all trapping events across every trajectory.

When $f = 0$, the normalized trapping time, $\tilde{T}_{\mathrm{trap}} = T_{\mathrm{trap}}/(NT)$, decreases monotonically with $\gamma$ [Fig.~\ref{fig2}(a)]. Surprisingly, when the flow oscillates with $f = 0.5$ and $A_\mathrm{f}=0.15$, $\tilde{T}_{\mathrm{trap}}$ first increases with $\gamma$ and then decreases, indicating that the transport of particles of intermediate lengths is the slowest.

\textit{Simulation}---As a phenomenological model, the experimental flow can be approximated by the stream function~\cite{Solomon1988,Torney2007,Khurana2011}: $\Phi = \pi^{-1} \sin\{\pi[x-A_{\mathrm{f}}\sin(2\pi f t)]\}\sin(\pi y)$. The flow velocity is $\mathbf{U} = (\partial \Phi/\partial y, -\partial \Phi/\partial x, 0)$ with $\Phi$ acting as a Hamiltonian. We construct the Poincar\'e sections of the model flow using the oscillation period $\tau_{\mathrm{os}} = 1/f$ as the sampling time interval. Quasi-periodic elliptic islands coexist with a chaotic region: for $f=0.5$, each vortex has a single period-1 island [Fig.~\ref{fig1}(f)]; for $f=1.0$, a large central period-1 island is surrounded by three smaller period-3 islands [Fig.~\ref{fig1}(h)]. The Poincar\'e sections agree well with the experimental flow visualizations [Figs.~\ref{fig1}(e) and \ref{fig1}(g)]. We characterize the local divergence of fluid particles by computing the finite-time Lyapunov exponent field~\cite{Haller15}, which also shows excellent agreement between the model and experimental flows~\cite{SM}.

We describe the dynamics of elongated particles as slender, rigid filaments entrained by the model flow. The filament centerline is $\mathbf{r}(s,t) = \mathbf{r}_c (t) + s\mathbf{p}$, where $\mathbf{r}_c$ is the center-of-mass (COM) position, $\mathbf{p}$ is the unit tangent vector, and the arclength $s\in[-\gamma/2, \gamma/2]$. Using the leading-order slender body theory~\cite{Cox1970} and enforcing zero-force and zero-torque conditions yield the translational and rotational motions~\cite{SM}: $\dot{\mathbf{r}}_c(t) = \gamma^{-1}\int_{-\gamma/2}^{\gamma/2}\mathbf{U}[\mathbf{r}(s,t)]\,ds$, and $\dot{\mathbf{p}}(t) = \Omega\mathbf{p}^{\perp}$, where $\mathbf{p}^{\perp}$ is the unit normal vector, and $\Omega$ is the angular velocity induced by the background flow, given by $\Omega = 12\gamma^{-3} \mathbf{p}^{\perp}\cdot \int_{-\gamma/2}^{\gamma/2} s\mathbf{U}[\mathbf{r}(s,t)]\,ds$. For short filaments ($\gamma \to 0$), $\dot{\mathbf{r}}_c$ reduces to the local flow velocity $\mathbf{U}[\mathbf{r}_c]$. As $\gamma$ increases, nonlocal entrainment of the background flow increases and couples the rotational and translational dynamics. This breaks the Hamiltonian structure of the system and drives filament trajectories to deviate from those of fluid particles. 

As shown in Fig.~\ref{fig2}(b), $\tilde{T}_{\mathrm{trap}}$ obtained from simulations with $f = 0.5$ and $T = 300\ \tau_0$ also varies non-monotonically with $\gamma$, in agreement with the experimental results [Fig.~\ref{fig2}(a)]. Since particles initially located in the elliptic islands tend to remain trapped for most $\gamma$, this transport anomaly is largely attributed to the behaviors of particles starting from the chaotic region. From simulations in which particles are released in the chaotic region, we find that the non-monotonic $\tilde{T}_{\mathrm{trap}}$-$\gamma$ relation is accompanied by frequent long-time trapping events in vortices for particles of certain lengths, peaking around $\gamma = 0.5$.
\begin{figure}[t]
\centering
\includegraphics[bb = 0 10 245 130, scale=1.0, draft=false]{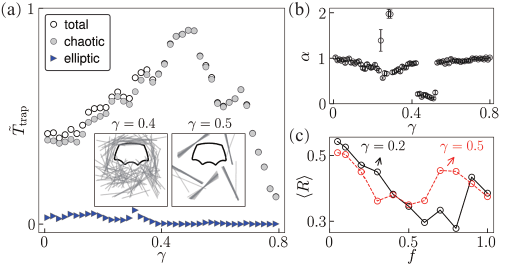}
\caption{(a) $\tilde{T}_{\mathrm{trap}}$ as a function of $\gamma$ for $f = 0.5$, $A_\mathrm{f} = 0.12$, and $T = 300\ \tau_0$. Particles are initially located in the chaotic region. Insets show the time lapse of particles at time interval $\tau_{\mathrm{os}}$ for $\gamma = 0.4$ and 0.5. (b) MSD exponent $\alpha$ as a function of $\gamma$ for $T = 10^4\ \tau_0$. (c) Average distance of particles to the nearest vortex centers $\langle R\rangle$ as a function of $f$.}
\label{fig3}
\end{figure}

By stroboscopically sampling particle trajectories at time interval $\tau_{\mathrm{os}}$ and examining their configurations on the flow Poincar\'e sections, we evaluate how chaotic and elliptic regions contribute to $\tilde{T}_{\mathrm{trap}}$ separately. As shown in Fig.~\ref{fig3}(a), for different $\gamma$, most trapping occurs in the chaotic region at $f = 0.5$. Particles with $\gamma$ around 0.5 do not meander like fluid particles; instead, they circulate on a period-5 orbit and remain trapped in vortices [Fig.~\ref{fig3}(a), right inset], producing the maximum in $\tilde{T}_{\mathrm{trap}}$. This trapping behavior arises from a resonance mechanism: a particle of the appropriate length reaches an orbit whose circulation period is commensurate with $\tau_{\mathrm{os}}$. For other lengths (e.g., $\gamma = 0.4$), particles appear more dispersed on the Poincar\'e section [Fig.~\ref{fig3}(a), left inset]. We compute the scaling exponent of the mean-squared displacement (MSD), $\langle \delta^2(\tau)\rangle \sim \tau^\alpha$, from long-time simulations with $T = 10^4\ \tau_0$. As shown in Fig.~\ref{fig3}(b), $\alpha$ drops abruptly nearly to zero around $\gamma = 0.5$. Interestingly, in a narrow range around $\gamma \approx 0.28$, a match between the particle transport time scale $W/|\dot{\mathbf{r}}_c|$ and half the oscillation period $\tau_{\mathrm{os}}/2$ leads to ballistic scaling with $\alpha \approx 2.0$, with particles taking directed, undulating steps along $\hat{\mathbf{x}}$~\cite{SM}.
\begin{figure*}[t]
\centering
\includegraphics[bb = 0 7 360 110, scale=1.37, draft=false]{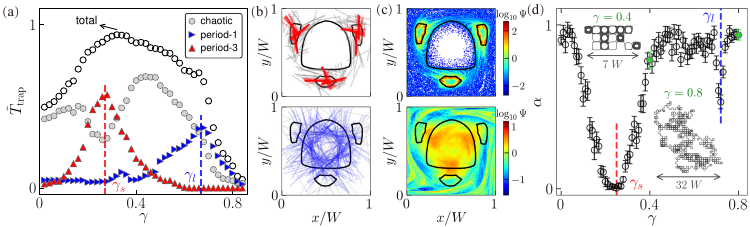}
\caption{Persistent trapping of particles (initially located in the chaotic region) in elliptic regions for $f = 1.0$ and $A_\mathrm{f} = 0.12$. Simulation duration $T = 300\ \tau_0$ in (a)--(c) and $T = 10^4\ \tau_0$ in (d). (a) $\tilde{T}_{\mathrm{trap}}$ as a function of $\gamma$. (b) Time lapse of particles at time interval $\tau_{\mathrm{os}}$ for $\gamma_s$ (top) and $\gamma_l$ (bottom). (c) $\Psi(x_c, y_c, t)$ for $\gamma_s$ (top) and $\gamma_l$ (bottom). (d) MSD exponent $\alpha$ as a function of $\gamma$. Insets show two typical trajectories at $\gamma = 0.4$ ($\alpha \approx 0.79$) and $\gamma = 0.8$ ($\alpha \approx 0.94$).}
\label{fig4}
\end{figure*}

\textit{Effect of elliptic islands}---In fact, similar non-monotonic $\tilde{T}_{\mathrm{trap}}$-$\gamma$ relations are observed when $f \gtrsim 0.4$. Importantly, $f$ controls whether particles starting from the chaotic region are influenced by the elliptic islands during transport. We compute the distance from particle COM positions to the nearest vortex centers, $\langle R \rangle$, averaged over time and across all trajectories. Figure~\ref{fig3}(c) shows that $\langle R \rangle$ decreases with $f$ and then fluctuates for $f\gtrsim 0.5$. This is especially pronounced for short particles. At low frequencies $f \sim 0.1$, particles near vortex boundaries alternate between neighboring vortices and execute normal diffusion. Meanwhile, they remain close to the boundaries and are rarely affected by the elliptic islands. As $f$ increases, particles shift toward the vortex centers, and long-time trapping events in the chaotic region start to appear for $f \gtrsim 0.4$, as exemplified by the case of $f=0.5$ [Fig.~\ref{fig3}(a)]. At higher frequencies $f\sim 1$, the COM speeds of particles are slow compared to the lateral oscillation of the vortices; over $\tau_{\mathrm{os}}$, they travel short distances, typically insufficient to escape the vortices in which they reside. In this regime, particles can reach the inner regions of the vortices and are more influenced by the elliptic islands.

At $f = 1.0$, $\tilde{T}_{\mathrm{trap}}$ also varies non-monotonically with $\gamma$ [Figure~\ref{fig4}(a)]. Unlike the case of $f = 0.5$, it results from trapping in both the elliptic and chaotic regions. Surprisingly, $\tilde{T}_{\mathrm{trap}}$ in the period-3 islands peaks at a smaller length $\gamma_s \approx 0.3$, while $\tilde{T}_{\mathrm{trap}}$ in the period-1 island peaks at a larger length $\gamma_l \approx 0.7$. For intermediate lengths $\gamma \approx 0.5$, trapping events mostly occur in the chaotic region. However, these events typically last less than $100\ \tau_0$ and are less persistent than those in the elliptic regions. At $\gamma_s$ and $\gamma_l$, particles overlap extensively with the corresponding islands [Fig.~\ref{fig4}(b)]. Accordingly, the instantaneous probability distribution of the COM positions, $\Psi(x_c,y_c,t)$, at $t = 300\ \tau_0$ is concentrated in the small and large islands for the short and long particles, respectively [Fig.~\ref{fig4}(c)]. These microscopic trapping behaviors are reflected in the long-time MSD exponent $\alpha$ [Fig.~\ref{fig4}(d)]. Short particles ($\gamma \ll 1$) exhibit normal diffusion with $\alpha \approx 1.0$. As $\gamma$ increases, two local minima, each surrounded by a subdiffusive regime with $\alpha < 1$, appear and coincide approximately with the peak positions of $\tilde{T}_{\mathrm{trap}}$ in the elliptic islands [Fig.~\ref{fig4}(a)]. For sufficiently long particles ($\gamma\sim 1$), normal diffusion reemerges. 
\begin{figure}[b]
\centering
\includegraphics[bb = 0 7 245 100, scale=1.0, draft=false]{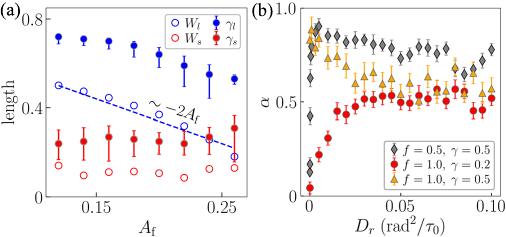}
\caption{(a) $W_l$, $\gamma_l$, $W_s$, and $\gamma_s$ versus $A_{\mathrm{f}}$ for $f = 1.0$. Error bars of $\gamma_l$ and $\gamma_s$ represent the widths of the minima. The dashed line is the best linear fit of $W_l$ with slope $-2$. (b) MSD scaling exponent $\alpha$ as a function of $D_r$ for $A_{\mathrm{f}}=0.12$.}
\label{fig5}
\end{figure}

Subdiffusion is further characterized by heavy-tailed distributions of trapping time~\cite{SM}. Compared to normal diffusion, subdiffusive dispersal is significantly slower, even when $\alpha$ is close to 1 [Fig.~\ref{fig4}(d) insets].

The subdiffusion at moderate ($f\sim 0.5$) and high ($f\sim 1$) frequencies shows distinct characteristics. At $f = 1.0$, only particles whose lengths match the sizes of elliptic islands undergo subdiffusive transport. This is due to broad spatial averaging of the nonuniform background flow across the islands along the particle length. Further evidence for this length-matching condition is found by varying $A_\mathrm{f}$. The size of the period-1 island $W_l$, defined as the square root of its area, decreases linearly with $A_{\mathrm{f}}$. Meanwhile, the larger local minimum in the $\alpha$-$\gamma$ relation $\gamma_l$ [Fig.~\ref{fig4}(d)] also decreases with $A_f$ [Fig.~\ref{fig5}(a)]. The size of the period-3 islands $W_s$ remains relatively constant, as does the corresponding $\gamma_s$. In addition, under the influence of the elliptic islands, the subdiffusion at $f = 1.0$ is more robust against random noise than that at $f=0.5$. We introduce Gaussian white noise $\xi(t)$ into the rotational equation: $\dot{\mathbf{p}}(t) = (\Omega + \sqrt{2D_r}\xi)\mathbf{p}^{\perp}$, where $D_r$ is the diffusion coefficient and $\xi$ satisfies $\langle \xi(t)\rangle = 0$ and $\langle \xi(t)\xi(t')\rangle = \delta(t-t')$. As shown in Fig.~\ref{fig5}(b), at $f = 0.5$, a small $D_r$ drastically raises $\alpha$ close to 1. In contrast, $\alpha$ at $f = 1.0$ is more resistant to the increase of $D_r$. For particles of intermediate lengths (e.g., $\gamma = 0.5$) that do not match either elliptic island, $\alpha$ at $f = 1.0$ even decreases with $D_r$, accompanied by a shift from transient trapping in the chaotic region to long-time trapping in the elliptic islands (not shown).

Finally, the particle dispersal patterns at spatial scales much larger than $W$ are shaped by $\gamma$. Short particles are primarily transported along the oscillation direction $\hat{\mathbf{x}}$. As $\gamma$ increases, the spread along the $\hat{\mathbf{y}}$ direction expands, forming elliptical dispersal patterns~\cite{SM}. As a hallmark of subdiffusion~\cite{Klafter2011}, the marginal distributions $\Psi(x_c,t)$ and $\Psi(y_c,t)$ for $\alpha < 1$ exhibit sharp cusps at the origin, in contrast to the Gaussian profiles seen in normal diffusion. 

\textit{Discussion}---Combining experiments and simulations, we demonstrate that elongated particles entrained by oscillatory cellular vortical flows exhibit rich transport dynamics---including subdiffusive, normal-diffusive, and ballistic transport---determined by the oscillation frequency $f$ and particle length $\gamma$. In particular, when $f \sim \tau_{\mathrm{os}}^{-1}$, particles with lengths matching the elliptic islands perform robust subdiffusion against random noise. This novel hydrodynamic length-selection effect resembles the action of mechanical mesh sieves and may be exploited to engineer particle diffusion and dispersal in fluid flows~\cite{Dorfman2013}. Active pointlike particles have been found to experience motility-dependent trapping in elliptic islands, but no subdiffusion occurs there~\cite{Khurana2011}. Finite-length effects are also important for particle dynamics in turbulence~\cite{Voth2017,Hu2021_2}. For instance, averaging over small-scale eddies reduces diffusion coefficients~\cite{Shin2005}. In this work, we focus on two-dimensional unsteady flows. Since entrapment in elliptic islands arises from spatial averaging of flow nonuniformity along particle length, similar length-selection effects may also emerge in three-dimensional chaotic flows with mixed advection properties~\cite{Speetjens2021}.

\bigskip

\begin{acknowledgments}
We thank Zhuang Su, Masao Doi, and Yi Peng for helpful conversations. S.H. and X.M. acknowledge support by National Natural Science Foundation
of China (NSFC Grant No. 22473005, 12247130, and 21961142020) and the Fundamental Research Funds for the Central Universities. J.Z. acknowledges support by NSFC (Grant No. 92252204).
\end{acknowledgments}

\bibliography{reference}

\end{document}


\title{Supplemental Material for \\
Anomalous Transport of Elongated Particles in Oscillatory Vortical Flows}

\author{Shiyuan Hu}
\thanks{These authors contributed equally to this work.}
\email[]{shiyuanhu@buaa.edu.cn}
\affiliation{School of Physics, Beihang Univerity, Beijing 100191 China}
\affiliation{CAS Key Laboratory of Theoretical Physics, Institute of Theoretical Physics, Chinese Academy of Sciences, Beijing 100190, China}
\author{Xiuyuan Yang}
\thanks{These authors contributed equally to this work.}
\affiliation{School of Physics, Beihang Univerity, Beijing 100191 China}
\author{Nan Luo}
\affiliation{Beijing National Laboratory for Condensed Matter Physics and Laboratory of Soft Matter Physics, Institute of Physics, Chinese Academy of Sciences, Beijing 100190, China}
\affiliation{School of Physical Sciences, University of Chinese Academy of Sciences, Beijing 100049, China}
\author{Jun Zhang}
\email[]{jun@cims.nyu.edu}
\affiliation{Applied Mathematics Lab, Courant Institute of Mathematical Sciences, and Department of Physics, New York University, New York, NY 10012, USA}
\affiliation{NYU-ECNU Institute of Physics at NYU Shanghai, Shanghai 200062, China}
\author{Xingkun Man}
\email[]{manxk@buaa.edu.cn}
\affiliation{School of Physics, Beihang Univerity, Beijing 100191 China}
\affiliation{Peng Huanwu Collaborative Center for Research and Education, Beihang University, Beijing 100191, China}

\date{\today}

\maketitle

\section{Supplemental videos}
Video 1: Oscillatory behavior of cellular vortical flows, showing the time evolution of the flow field, center positions of vortex (red dot) and magnet (blue dot), and the finite-time Lyapunov exponent (FTLE) field. The flow oscillation frequency $f = 1.0$ and amplitude $A_\mathrm{f} = 0.12$. 

Video 2: Typical trajectories of particles in steady flows. The particle length $L = 2$ mm and 6 mm. 

Video 3: Flow visualizations of dynamical structures in oscillatory flows. The chaotic region is marked by green dye. The oscillation amplitude $A_{\mathrm{f}} = 0.12$, and two frequencies are recorded, $f = 0.5$ and 1.0. Both continuous and stroboscopic recordings are shown.

\section{Scaling relation of flow oscillation amplitude}
\begin{figure*}[h]
\centering
\includegraphics[bb = 0 5 250 120, scale=1.0, draft=false]{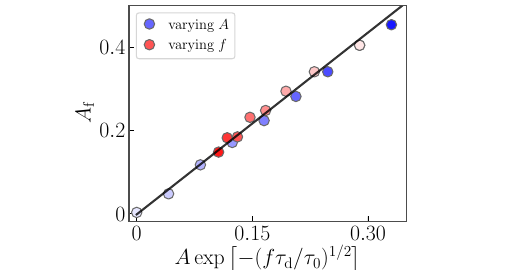}
\caption{Flow oscillation amplitude $A_{\mathrm{f}}$ versus $A \exp [-(f\tau_{\mathrm{d}}/\tau_0)^{1/2}]$, where $A$ is the oscillation amplitude of the magnets, $\tau_\mathrm{d} = d^2/\nu$, and $\tau_0 = W/U_0$. Lengths are scaled by the vortex size $W$ and frequency by $\tau_0^{-1}$. Blue circles show data points with varying $A$, and red circles show those with varying $f$. The magnitudes of both $A$ and $f$ increase from light to dark colors. The solid line is the best linear fit.}
\label{figS1}
\end{figure*}
In the classical solution for flow induced by an oscillating plane boundary, the flow velocity decays exponentially as $\exp[-(f_\mathrm{b} h^2/\nu)^{1/2}]$, where $f_\mathrm{b}$ is the oscillation frequency of the boundary and $h$ is the distance from the boundary. The amplitude $A_{\mathrm{f}}$ follows the same scaling [Fig.~\ref{figS1}]. Therefore, the flow field in the upper flow layer in our experiments is primarily driven by viscous stress across the fluid depth.

\section{Finite-time Lyapunov exponent field}
\begin{figure*}[h]
\centering
\includegraphics[bb = 0 0 245 105, scale=1.0, draft=false]{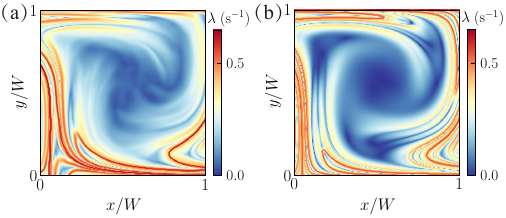}
\caption{Finite-time Lyapunov exponent field of the (a) experimental flow and (b) model flow. The time interval $\Delta t = 4\ \tau_0$. The flow oscillation amplitude $A_{\mathrm{f}} = 0.12$ and frequency $f = 1.0$.}
\label{figS2}
\end{figure*}
In time-dependent flows, unlike instantaneous streamlines, which can quickly diverge from the actual trajectories of tracer particles, the finite-time Lyapunov exponent (FTLE) field is more indicative of particle trajectories and transport behavior~\cite{Haller15}. We define the flow map as $\textbf{x} = \Phi(\textbf{x}_0,t_0,\Delta t)$, which maps a tracer particle initially located at $\textbf{x}_0$ at time $t_0$ to $\textbf{x}$ after a time interval $\Delta t$. To compute $\Phi$, the position of the tracer particle is evolved in time by integrating the velocity field. We then compute the right Cauchy-Green strain tensor, $C_{ij} = \left(\partial \Phi_k/\partial x_{0,i}\right) \left(\partial \Phi_k/\partial x_{0,j}\right)$, where summation is taken over the repeated index $k = 1,2$. The gradients $\partial \Phi_k/\partial x_{0,i}$ are computed numerically using a grid-based method: a set of tracer particles is initialized on a uniform grid; after $\Phi$ is obtained, the gradients are evaluated using a second-order finite-difference scheme. We next calculate the largest eigenvalue of Cauchy-Green tensor, $\lambda (\textbf{x}_0,t_0,\Delta t)$. Finally, the FTLE is calculated as
\begin{equation}\label{FTLE}
\sigma(\textbf{x}_0, t_0, \Delta t) = \frac{1}{|\Delta t|}\ln \sqrt{\lambda(\textbf{x}_0, t_0, \Delta t)}.
\end{equation}
As shown in Fig.~\ref{figS2}, the FTLE fields of the experimental flow and the model flow agree well. The regions between two vortices have large values of FTLE, indicating diverging trajectories. Near the vortex centers, the values of FTLE are small, and fluid particles remain trapped.

\section{Simulation model}
Consider a rigid filament of length $L$ entrained by a background flow field $\mathbf{U}$. The filament centerline position is given by $\mathbf{r}(s,t) = \mathbf{r}_c + s\mathbf{p}$, where the signed arc length $s\in [-L/2, L/2]$, and the filament velocity $\mathbf{r}_t = \dot{\mathbf{r}}_c + s\dot{\mathbf{p}}$. From the leading-order slender-body theory, the filament velocity $\mathbf{r}_t$ is related to the hydrodynamic force density $\mathbf{f}$ through
\begin{equation}
8\pi\mu (\mathbf{r}_t - \mathbf{U}[\mathbf{r}]) = -c(\mathbf{I}+\mathbf{r}_s\mathbf{r}_s)\cdot \mathbf{f},
\end{equation}
where $\mu$ is the dynamic viscosity of the fluid and the slenderness parameter $c = |\ln(\epsilon^2 e)|$ with $\epsilon$ the filament aspect ratio. Since the filament is passively advected by the flow, the total force and torque are zero,
\begin{equation}
\int_{-L/2}^{L/2} \mathbf{f}\,ds = 0,\quad \int_{-L/2}^{L/2}s\mathbf{p}\times \mathbf{f}\,ds = 0.
\end{equation}
Enforcing these two constraints, we obtain
\begin{gather}
\dot{\mathbf{r}}_c = \frac{1}{L}\int_{-L/2}^{L/2}\mathbf{U}[\mathbf{r}(s)]\,ds, \label{com_velo}\\
\dot{\mathbf{p}} = \Omega\mathbf{p}^{\perp}= \frac{12}{L^3} (\mathbf{I}-\mathbf{p}\mathbf{p})\cdot \int_{-L/2}^{L/2}s\mathbf{U}[\mathbf{r}(s)]\,ds, \label{p_dot}
\end{gather}
where the angular velocity $\Omega$ induced by the fluid flow is given by
\begin{equation}\label{theta_dot}
\Omega = \frac{12}{L^3}\mathbf{p}^{\perp}\cdot \int_{-L/2}^{L/2}s\mathbf{U}[\mathbf{r}(s)]\,ds.
\end{equation}
Equations~(\ref{com_velo}) and (\ref{p_dot}) are evolved using a fourth-order Runge-Kutta scheme. 

In the presence of rotational diffusion, Eq.~(\ref{theta_dot}) is modified as 
\begin{equation}
\Omega = \frac{12}{L^3}\mathbf{p}^{\perp}\cdot \int_{-L/2}^{L/2}s\mathbf{U}[\mathbf{r}]\,ds + \sqrt{2D_r} \xi,
\end{equation}
where $D_r$ is the diffusion constant, and the white noise $\xi$ satisfies the statistics
\begin{gather}
\langle \xi (t) \rangle = 0,\\
\langle \xi(t) \xi(t') \rangle = \delta (t-t').
\end{gather}
Equations~(\ref{com_velo}) and (\ref{p_dot}) are made nondimensional in the main text by scaling lengths with $W$, time with $\tau_0$, and velocity with $U_0$.

\section{Ballistic trajectories}
\begin{figure*}[h]
\centering
\includegraphics[bb = 0 5 360 80, scale=1.2, draft=false]{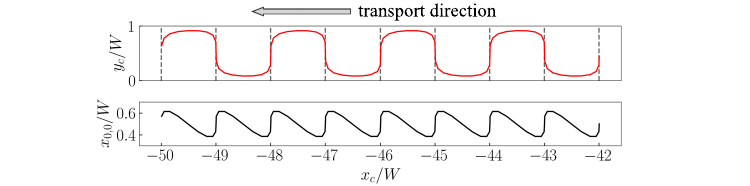}
\caption{Ballistic motion of particles. The oscillation amplitude $A_{\mathrm{f}}=0.12$ and frequency $f=0.5$. The particle length $\gamma = 0.28$. Top panel: a segment of a typical ballistic trajectory. The particle is moving toward the $-\hat{\mathbf{x}}$ direction. Dashed lines indicate vortex boundaries. Bottom panel: the $x$ coordinate of one of the vortex centers $x_{0,0}$ plotted as a function of the particle COM position $x_c$. In the model flow, the vortex centers are located at $\mathbf{x}_{m,n} = (1/2+m, 1/2+n)$ for $m$, $n$ integers.}
\label{figS3}
\end{figure*}
At moderate oscillation frequency $f = 0.5$, particles of certain lengths can reach undulating ballistic trajectories moving along the oscillation direction $\hat{\mathbf{x}}$. As shown in Fig.~\ref{figS3}, each time the particle moves in the $-\hat{\mathbf{x}}$ direction, the vortices shift toward the $+\hat{\mathbf{x}}$ direction, enabling the particle to cross the vortex boundaries and enter the next vortex. This behavior results from a match between the particle characteristic transport time $W/\dot{\mathbf{r}}_c$ and half the oscillation period $\tau_{\mathrm{os}}/2$. 

\section{Macroscopic Dispersal patterns}
\begin{figure*}[h]
\centering
\includegraphics[bb = 0 0 365 195, scale=1.35, draft=false]{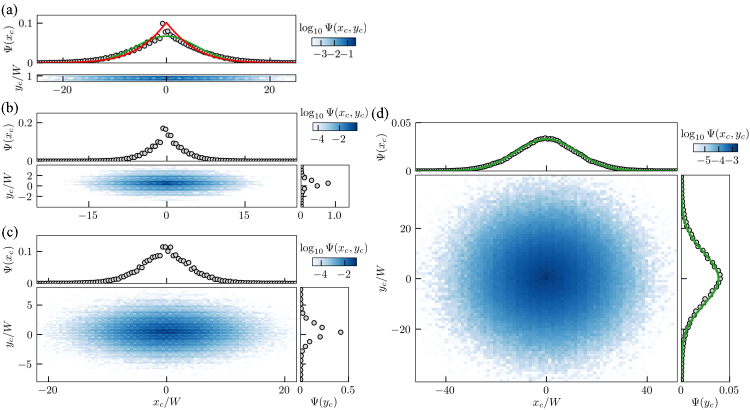}
\caption{Dispersal patterns for different particle lengths. (a) $\gamma = 0.1$, (b) $\gamma = 0.4$, (c) $\gamma = 0.5$, and (d) $\gamma = 0.8$. The marginal distributions $\Psi(x_c, t)$ and $\Psi(y_c, t)$ are shown on the upper and right axes, respectively. Green curves represent the best fit Gaussian distributions. The red curve in (a) plots the analytic distribution [Eq.~(\ref{sub_distribution})] for one-dimensional subdiffusive transport without fitting parameters.}
\label{figS4}
\end{figure*}

The transport of particles in cellular flows can be viewed as a random-walk process. Consider a one-dimensional random walk model, where the random walker instantaneously jumps to a new position by a step of length $l$ and then waits for a duration $\tau$ before the next jump. In our system, the waiting time corresponds to the trapping time of a particle in a particular vortex. The direction of each jump is chosen at random. The probability distribution of step length $p(l)$ is assumed to be Gaussian with variance $\sigma^2$, and the waiting time $\tau$ is drawn from a power-law distribution, $\phi(\tau) \sim \tau_{\mathrm{w}}^\beta \tau^{-(1+\beta)}$, where $\tau_{\mathrm{w}}$ is a characteristic time scale. For $0 < \beta < 1$, the random walk is subdiffusive with the MSD scaling exponent $\alpha = \beta$. Let $\Psi(x,t)$ denote the probability density of finding the random walker at position $x$ at time $t$. In Fourier-Laplace space, the probability density has an analytical form, given by~\cite{Klafter2011}
\begin{equation}\label{sub_distribution}
\Psi(k,s) = \int_{-\infty}^{\infty}\int_0^{\infty} \Psi(x,t) e^{ikx} e^{-st}\,dx\,dt = \frac{\tau_{\mathrm{w}}^\beta s^{\beta-1}}{k^2\sigma^2/2 + \tau_{\mathrm{w}}^\beta s^\beta}.
\end{equation}
The real-space probability density $\Psi(x,t)$ can be obtained by numerically inverting the Fourier and Laplace transforms. 

Figure~\ref{figS4} shows the dispersal patterns of $5\times 10^5$ particles at simulation time $t = 500\ \tau_0$. Particles are initially distributed in the chaotic region of the unit cell centered at the origin. For $\gamma = 0.1$ [Fig.~\ref{figS4}(a)], particles are primarily transported along the oscillation direction $\hat{\mathbf{x}}$, and the random walk is one-dimensional. By taking $\tau_{\mathrm{w}} = \tau_0$, $\sigma = W$, and $\beta = \alpha \approx 0.55$, Eq.~(\ref{sub_distribution}) agrees well with the marginal distribution $\Psi(x_c, t)$ obtained from particle simulations without fitting parameters. For $\gamma = 0.4$ and 0.5 [Figs.~\ref{figS4}(b) and \ref{figS4}(c)], the dispersal in the $\hat{\mathbf{y}}$ direction increases, but the overall dispersal pattern remains anisotropic. The sharp cusps at the origins of the marginal distributions $\Psi(x_c, t)$ and $\Psi(y_c, t)$ persist. For $\gamma = 0.8$ [Fig.~\ref{figS4}(d)], the transport becomes normal diffusive, and both $\Psi(x_c, t)$ and $\Psi(y_c, t)$ fit well to Gaussian distributions.
\begin{figure*}[b]
\centering
\includegraphics[bb = 0 0 365 155, scale=0.7, draft=false]{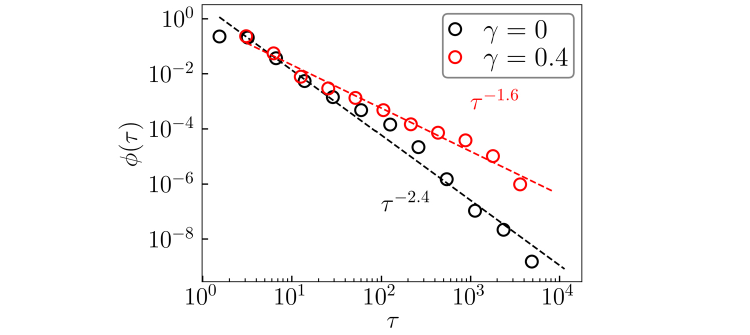}
\caption{Distributions of trapping time for $\gamma = 0$ and $\gamma = 0.4$. The oscillation amplitude $A_\mathrm{f} = 0.12$ and frequency $f = 1.0$.}
\label{figS5}
\end{figure*}

\section{Distributions of trapping time}
The distributions of trapping time $\phi(\tau)$ for $A_\mathrm{f} = 0.12$ and $f = 1.0$ are shown in Fig.~\ref{figS5}. Straight lines represent the best fits of power-law distributions: $\phi (\tau) \sim \tau^{-(1+\beta)}$. For $\gamma = 0$ (fluid tracers), $\beta \approx 1.4$, and for $\gamma = 0.4$, $\beta \approx 0.6$. In random walk theory~\cite{Klafter2011}, $\beta < 1$ leads to subdiffusion, whereas $\beta > 1$ leads to normal diffusion.

\bibliography{reference}